\documentstyle[twocolumn,prl,aps,epsfig,amssymb]{revtex}

\def\beqn{\begin{eqnarray}}
\def\eeqn{\end{eqnarray}}
\relax
\newcommand{\ba}[1]{\begin{array}{#1}}
\def\ea{\end{array}}

\def\beq{\begin{equation}}
\def\eeq{\end{equation}}
\def\bea{\begin{array}}
\def\eea{\end{array}}

\def\to{\rightarrow}

\def\dis{\displaystyle}
\def\f{\frac}
\def\[{\left[}
\def\]{\right]}
\def\({\left(}
\def\){\right)}

\def\gm2{{g_\mu -2}}
\def\hl{\hline}
\def\wH{\widetilde{H}}

\def\ep{{\epsilon}}

\def\sm0{{\widetilde{m}_0}}

\def\cut{{\Lambda}}

\def\ov{\overline}

\def\U1em{{U(1)_{\rm em}}}
\def\to{\rightarrow}

\def\sq2{\sqrt{2}}

\def\tanb{\tan\hspace*{-1mm}\beta}

\def\End{\end{document}}

\begin{document}                                                              
\draft

\twocolumn[\hsize\textwidth\columnwidth\hsize\csname
@twocolumnfalse\endcsname
     
\setcounter{footnote}{0}

\title{  
Neutrino-Lepton Masses,  Zee Scalars and Muon $g - 2$
}  
\author{
{\sc Duane A. Dicus}\,$^1$,~~
{\sc Hong-Jian He}\,$^1$,~~  
{\sc John N. Ng}\,$^{2,3}$
}
\address{
\vspace*{2mm}
$^1$The University of Texas at Austin, Austin, Texas 78712, USA\\
$^2$TRIUMF, 4004 Wesbrook Mall, Vancouver, BC V6T 2A3, Canada\\
$^3$National Center for Theoretical Science, 
    P.O. Box 2-131, Hsinchu, Taiwan
}
%
\maketitle
\begin{abstract}
\hspace*{-0.2cm}
Evidence for neutrino oscillations is pointing to
the existence of tiny but finite neutrino masses.
Such masses may be naturally
generated via radiative corrections in models such as the Zee model
where a singlet Zee-scalar plays a key role.
We minimally extend the Zee model by including a right-handed
singlet neutrino $\nu_R$. The radiative Zee-mechanism 
can be protected by a simple $U(1)_X$  symmetry 
involving only the $\nu_R$ and a Zee-scalar. 
We further construct a class of models with a single horizontal $U(1)_{\rm FN}$
(\`{a} la Froggatt-Nielsen) such that the mass patterns of the
neutrinos and leptons are naturally explained.
We then analyze the muon anomalous magnetic moment ($\gm2$) and the flavor
changing $\mu\to e\gamma$ decay.
The $\nu_R$ interaction in our minimal extension is found to
induce the BNL $\gm2$ anomaly, with a light charged Zee-scalar of mass
$100-300$\,GeV.
\\
{PACS numbers:\,14.60.Pq,\,12.15.Ff,\,13.15.$+$g,\,13.40.Em\,
\hfill   
[ hep-ph/0103126 ] 
} 
\end{abstract}
\vskip 1pc]


\setcounter{footnote}{0}
\renewcommand{\thefootnote}{\arabic{footnote}}

The standard model (SM) of the electroweak and strong interactions has
to be extended, in light of the existing     
neutrino oscillation data\,\cite{nu-data} which provides
strong evidence for tiny but finite neutrino masses.
The squared neutrino-mass differences are found to
have two distinct ranges, 
$10^{-11}\,{\rm eV}^2 \leq \delta m_{\odot}^2 \leq 10^{-5}\,{\rm eV}^2$ 
and 
$\delta m_{\rm atm}^2\simeq 10^{-3}\,{\rm eV}^2$,
in order to interpret the solar and atmospheric neutrino anomalies, respectively.
The very small neutrino masses may be explained by invoking
the seesaw mechanism\,\cite{seesaw} where a heavy right-handed Majorana
singlet neutrino is introduced at the grand unification scale
($10^{10-16}$\,GeV).
On the other hand, the radiative mechanism 
for neutrino mass generation,
such as advocated in the Zee model\,\cite{zee}, 
provides an important alternative, where the relevant new physics is 
expected to show up at or near the weak scale. 
Such low scale models, besides being able to explain some
features of neutrino oscillations such as the bi-maximal mixing\,\cite{Jar}, 
are clearly of phenomenological relevance. 
The existing precision data have put some nontrivial constraints 
on the models\,\cite{Jar,YuS,Ng} and further tests will be available
at forthcoming collider experiments.

The minimal Zee model\,\cite{zee} contains the
three active left-handed neutrinos
of the SM and a bilepton singlet Zee-scalar which plays a key role
for radiative generation of their Majorana masses.
There is no underlying reason that forbids the existence of light 
right-handed singlet neutrinos ($\nu_R$), as 
$\nu_R$ can also be naturally contained in various extensions of the SM
(such as the models with left-right symmetry or $SO(10)$).
The introduction of $\nu_R$ thus provides the simplest possible
extension of Zee model. However, 
a simple embedding of  $\nu_R$ results in the loss of the prediction
of neutrino masses as they are rendered arbitrary by the tree-level 
Dirac mass terms\,\cite{Wolf,Yanagida,YuS,Ng}.
In this work, we first build a class of minimally extended Zee models including
$\nu_R$ (called Type-I) and invoke
a simple $U(1)_X$ symmetry (or its discrete subgroup $Z_n$) 
to effectively protect the radiative Zee-mechanism by forbidding the
mixings between the $\nu_R$ and the active neutrinos. 
Such an extension is nontrivial since the successful 
embedding of a singlet $\nu_R$  requires the addition of 
a second singlet Zee-scalar in the minimal extension. 
Next we note that the original Zee model neither predicts 
the size of the Zee-scalar Yukawa couplings
nor provides any insight on generating
the lepton masses and their hierarchy. 
Though our simplest $U(1)_X$ in Type-I models
protects the radiative Zee-mechanism, we can use this
{\it same} $U(1)_X$ group 
as a horizontal symmetry involving both neutrinos and leptons 
(called Type-II), and thus explain the mass patterns of the
neutrinos and leptons in a natural way,
\`{a} la Froggatt-Nielsen\,\cite{FN}.  
In both Type-I and -II models, the $\nu_R$ can interact with
the right-handed muon and Zee-scalar with a
natural $O(1)$ Yukawa coupling, which has striking phenomenological 
consequences.

Finally, we apply the Type-I and -II models to analyze the
Zee-scalar-induced contributions to the muon anomalous magnetic moment
$\gm2$ and the lepton-flavor-violation decay
$\mu\to e\gamma$\,. We find that the recent  
BNL $\gm2$ anomaly\,\cite{BNL}
can be explained with a light charged Zee-scalar of mass around
$100-300$\,GeV.
Our models also have the $\mu\to e\gamma$ decay branching ratio around or below
the current experimental limit.

\vspace*{0.2cm}
\noindent
\underline{\it Minimal Extension of the Zee Model}\\[1mm] 
\underline{\it with a Right-handed Singlet Neutrino}~\\[-2.5mm]
  
The minimal Zee model\,\cite{zee} introduces 
one extra singlet charged scalar ($S_1^\pm$) together with the usual 
two-Higgs-doublet sector. By assuming no right-handed $\nu_R$,
as in the SM,  this scalar only interacts
with left-handed neutrinos and leptons.
Thus, the Zee model contains the following additional Lagrangian,
\vspace*{-2mm}
\beqn
&
\hspace*{-3mm}
\Delta {\cal L}_1 \!=\!\!
\dis\sum_{j,j'} \f{f_{jj'}}{2}
\ep_{ab} \ov{L^c_a}_{j}L_{bj'} S_1^+
\!+m_3\ep_{ab}\wH_1^a{H}_2^b S_1^+ \!+\! {\rm h.c.} 
\nonumber 
&\\[0.25mm]
&\hspace*{-13.5mm}
=\!\[f_{12} \(\ov{\nu_e^c}  \mu_L \!-\ov{\nu_\mu^c}  e_L \) +
  f_{13} \(\ov{\nu_e^c}  \tau_L\!-\ov{\nu_\tau^c} e_L \) +\right. 
& 
\label{eq:ZeeL}
\\[1mm]
&
\hspace*{-1mm}
\left.
~~~~f_{23}\(\ov{\nu_\mu^c}\tau_L\!-\!\ov{\nu_\tau^c}\mu_L\)
\!+m_3(H_1^{0\ast}H_2^- \!-\! H_1^-H_2^0)\] S_1^+\!+\!{\rm h.c.} 
&
\nonumber
\eeqn
where $\ell_j,\ell_{j'} \in (e,\,\mu,\,\tau)$ and
$L_j = (\nu_{j},\,\ell_{jL})^T$ is the left-handed doublet
of the jth family.
The $(H_1,\,H_2)$ are the usual two-Higgs-doublets with hypercharge
$(1/2,\,-1/2)$, where 
$H_1=(-H_1^+,\,H_1^0)^T$,
$H_2=(H_2^0,\,H_2^-)^T$, and $\wH_j=i\tau_2 H_j^\ast$. 
The Yukawa sector can conserve total lepton number by assigning 
to $S_1^\pm$ the
lepton numbers $\mp 2$. Thus, the total lepton number is only softly
violated by the dimension-3 trilinear Higgs operator in Eq.\,(\ref{eq:ZeeL}).
As such, the small Majorana neutrino masses are radiatively generated
at one-loop and are automatically finite.

We minimally extend the Zee model by including a single
right-handed Dirac neutrino $\nu_R$ with following Yukawa interactions,
\beq
\Delta {\cal L}_2 ~=~ 
\[
 f_1\, \ov{\nu_{R}^c}e_R 
+f_2\, \ov{\nu_{R}^c}\mu_R
+f_3\, \ov{\nu_{R}^c}\tau_R
\] S_2^+ + {\rm h.c.} 
\label{eq:ZeeR}
\eeq
where $S_2^\pm$ is a second singlet Zee-scalar. The nontrivial issue with 
embedding $\nu_R$ is to avoid arbitrary tree-level Dirac mass terms 
generated by the Yukawa interactions 
$\ov{L_j}H_2\nu_R$ and
$\ov{L_j}\wH_1\nu_R$ (which mix $\nu_j$ and $\nu_R$), so that
the predictive power of the radiative Zee-mechanism can be effectively 
protected. We achieve this goal by noting that the Yukawa sector
(\ref{eq:ZeeR}) of $\nu_R$ possesses a global $U(1)_X$ symmetry, which
can properly forbid the neutrino Dirac mass terms once a Zee-scalar
$S_2^\pm$ is included together with $\nu_R$. 
It can be shown, by assigning the most general $U(1)_X$ 
quantum numbers for the Zee-model with $\nu_R$, 
that the unwanted tree-level neutrino Dirac masses cannot be removed
without $S_2^\pm$.
We define our simplest Type-I models with  
$U(1)_X$ in Table\,1, where only $\nu_R$ and $S_2^\pm$ carry $U(1)_X$ charges
while all other fields are singlets of $U(1)_X$. Hence, the Type-I extension
gives a truely minimal embedding of $\nu_R$ into the Zee model. 

\vspace*{2mm}
\noindent
{\small 
Table\,1. Quantum number assignments for Type-I and -II models. 
The hypercharge is defined as ${Y=Q - I_3}$\,. 
}
\begin{center}
\setlength{\tabcolsep}{0.2pc}
\begin{tabular}{c|cccccccc}
\hline\hline
&&&&&&&&\\[-2.5mm]
     &  $L_j$  & $\ell_{jR}$  &  \,$\nu_R$\, 
                              & $H_1$  &  $H_2$  & $~S_1^+$ &  $~S_2^+$ & $~S^0$~           
\, \\ [1.5mm] 
\hline
&&&&&&&&\\[-2.5mm]
$U(1)_Y$   & \,$-1/2$\,  &  $-1$  &  $0$  & \,$1/2$\,  & \,$-1/2$\,  & $1$  & $1$  & $0$~  \\[1.5mm]
\hline
&&&&&&&&\\[-2.5mm]
$U(1)_X$   & $0$     &  $0$   &  $x$  & $0$    & $0$     & $0$  & $-x$  & $-$~ \\[1.5mm]
\hline
&&&&&&&&\\[-2.5mm]
$U(1)^a_{\rm FN}$       
           & $0$     &  $y_j$ &  $x'$ & $0$    & $z$     & $-z$ & $-x\!-\!y$ & $-1$~ \\[1.5mm]
\hline
&&&&&&&&\\[-2.5mm]
\,$U(1)^b_{\rm FN}$\,       
           & $u_j$   &  $y_j$ &  $x'$ & $0$    & $z$     & $-z$ & $-x\!-\!y$ & $-1$~ \\[1.5mm]
\hline\hline                 
\end{tabular}
\end{center}

\vspace*{2mm}
Table\,2 classifies all (dis-)allowed operators of Type-I up to dimension-4. 
It shows that, as long as $x\neq 0$, the radiative Zee-mechanism is protected
and the $\nu_R$ remains massless. Such a massless $\nu_R$ does not
contribute to the invisible $Z$-width as it carries no weak charge. 
A special case of our Type-I is to consider its discrete subgroup
$Z_4$ under which $\nu_R$ and $S_2^\pm$ transform as,
$\nu_R\to i\nu_R\,,~ S_2^\pm \to \mp i S_2^\pm\,,$
while all other fields remain invariant.
Other non-minimal variations of our Type-I can be easily constructed.

\vspace*{2.5mm}
\noindent
{\small 
Table\,2. Summary of $U(1)$ charges carried by the effective operators 
in Type-I and -II models.
}
\begin{center}
\vspace*{0mm}
\setlength{\tabcolsep}{0.6pc}
\begin{tabular}{c|ccc}
\hline\hline
&&&\\[-2.5mm]
~Operators                       & $U(1)_X$  & $U(1)^a_{\rm FN}$  & $U(1)^b_{\rm FN}$\\[1mm] 
\hline
&&&\\[-3.2mm]
~$\ov{L}_{j'} H_1\ell_{jR}$      & $0$       & $y_j$         & $y_j-u_{j'}$       \\ [.5mm] 
\hl
&&&\\[-3.2mm]
~$\ov{L}_{j'}\wH_2\ell_{jR}$     & $0$       & $y_j\!-\!z$   & $y_j\!-\!z-\!u_{j'}$     \\ [.5mm] 
\hl
&&&\\[-3.2mm]
~$\ov{L}_{j'}\wH_1\nu_R$         & $x$       & $x'$      & $x'\!-\!u_{j'}$         \\ [.5mm] 
\hl
&&&\\[-3.2mm]
~$\ov{L}_{j'} H_2\nu_R$          & $x$       & $x'\!+z\!$ & $x'\!+\!z\!-\!u_{j'}$  \\ [.5mm]   
\hl
&&&\\[-3.2mm]
~$\ov{\nu_{j'}^c}\ell_{jL}S_1^+$ & $0$       & $-z$      & $u_j\!+\!u_{j'}\!-\!z$  \\ [.5mm] 
\hl
&&&\\[-3.2mm]
~$\ov{\nu_{j'}^c}\ell_{jL}S_2^+$ & $-x$      & $-x\!-\!y$ & $u_j\!+\!u_{j'}\!-x\!-\!y$ \\ [.5mm]  
\hl
&&&\\[-3.2mm]           
~$\ov{\nu_R^c}\ell_{jR}S_1^+$    & $x$       & $x'\!+\!y_j\!-\!z$ & $x'\!+\!y_j\!-\!z$ \\ [.5mm]  
\hl
&&&\\[-3.2mm]
~$\ov{\nu_R^c}\ell_{jR}S_2^+$    & $0$       & $x'\!+\!y_j\!-\!x\!-\!y$ & $x'\!+\!y_j\!-\!x\!-\!y$ 
                                                                                      \\ [.5mm]
\hl
&&&\\[-3.2mm]
~$\wH_1 H_2 S_1^+$               & $0$       & $0$       & $0$                \\ [.5mm]
\hl
&&&\\[-3.2mm]
~$\wH_1 H_2 S_2^+$               & $-x$      & $z\!-\!x\!-\!y$   & $z\!-\!x\!-\!y$ \\ [.5mm]
\hl
&&&\\[-3.2mm]
~$\nu_R^T\nu_R$                  & $2x$      & $2x'$     & $2x'$              \\ [.5mm]
\hl
&&&\\[-3.2mm]
~$S_1^+ S_2^-$                   & $x$       & $x\!+\!y\!-\!z$   & $x\!+\!y\!-\!z$ \\ [.5mm]
\hl
&&&\\[-3.2mm]
~$\ep_{ab}H_1^aH_2^b$            & $0$       & $z$       & $z$                \\ [.5mm]
\hline\hline
\end{tabular}
\end{center}

\vspace*{0.4cm}
\noindent
\underline{\it Neutrino Oscillations, Lepton Masses and }\\[1mm] 
\underline{\it Horizontal $U(1)_{\rm FN}$ Symmetry}\\[-2.5mm]

While the above Type-I models give the most economic embedding of
$\nu_R$ with all the good features of the 
original Zee-model retained, they do not 
provide any insight on two important issues: (i) There is no theory
prediction on the size of the
Zee-scalar Yukawa couplings $f_{jj'}$ in Eq.\,(1), but
the neutrino oscillation data requires
the following hierarchy\,\cite{Jar}:
\vspace*{-1.8mm}
\beq
\ba{l}
\dis \f{f_{12}}{f_{13}} \simeq \f{m_\tau^2}{m_\mu^2}\simeq 3\times 10^2,~~~
\dis \f{f_{13}}{f_{23}} \simeq \f{\sqrt{2}\delta m_{\rm atm}^2}
                                 {\delta m_{\odot}^2}\simeq 10^2 ~{\rm or}~ 10^7,
\\[-4.5mm]
\ea
\label{eq:nuCond}
\eeq
where ${f_{13}}/{f_{23}} \simeq 10^2\,(10^7)$ corresponds to the MSW
large angle solution (vacuum oscillation solution).  
(ii) The small lepton masses and their large hierarchy are not understood.
Our goal is to construct this {\it same} $U(1)$ group as a 
horizontal symmetry involving all the leptons so that these two issues
can be naturally explained  \`{a} la Froggatt-Nielsen (FN)\,\cite{FN}.
[This $U(1)$ will be called $U(1)_{\rm FN}$.]
The basic idea is to consider a horizontal $U(1)_{\rm FN}$
spontaneously broken by the vacuum expectation value 
$\langle S^0\rangle$ of a singlet scalar $S^0$. We can assign
$U(1)_{\rm FN}$ charges for relevant fields such that different mass terms are
suppressed by different powers of $\ep \equiv\langle S^0\rangle/\cut$ where
$\cut$ is the scale at which the $U(1)_{\rm FN}$ breaking is mediated
to the light fermions. 
For instance, a low energy effective operator 
carrying a net $U(1)_{\rm FN}$ charge $q$  (either $\geq0$ or $<0$)
will be suppressed by $\ep^{|q|}$.
Though all mass terms are now allowed in the effective 
theory, we will build a class of  FN-type models (called Type-II) in 
which the arbitrary tree-level neutrino Dirac-mass terms are 
suppressed to a level much below the one-loop radiative Zee-masses, and thus 
the predictive power of the Zee-mechanism remains.
The role of the FN-scalar $S^0$  is to provide the spontaneous 
$U(1)_{\rm FN}$ breaking and generate the 
relevant $U(1)_{\rm FN}$-invariant effective
operators that will give the desired neutrino Yukawa couplings and 
lepton masses at the weak scale. 
The heavy $S^0$ will be eventually integrated out from the low energy theory
and our relevant particle spectrum of Type-II is the {\it same} as Type-I.

We provide two typical Type-II constructions, 
called Type-IIa and -IIb, respectively.
The Type-IIa is the simplest extension of Type-I by further involving only the 
right-handed weak-singlet leptons in the $U(1)_{\rm FN}$ (cf. Table\,1).
In the Type-IIb models, we further assign each lepton doublet $L_j$ a 
charge $u_j$.
So, the lepton masses are determined by $\ell_{jR}$ charges in Type-IIa, while
Type-IIb determines these masses by the charges of both $\ell_{jR}$ and
$L_j$. The low energy effective operators up to dimension-4 
(with the heavy $S^0$ integrated out) are classified in Table\,2, from which we
derive the general conditions for protecting the Zee-mechanism in Type-II,
\beq
10 > |x'| \sim |x| \gg 1,~~~{\rm and}~~~ |x-u_j|,\, |x+y|  \gg  1,  
\label{eq:ZeeCond}
\eeq
with \,$xx'>0$\, and \,$|y|,\,|z|\sim O(1)$.\,
For the explicit analysis below,
we choose a typical value of the suppression factor $\ep \simeq 0.1$.
Thus, choosing leptons in mass-eigenbasis, we write
their mass ratios as,
\vspace*{-1mm}
\beq 
m_e\,:\,m_\mu\,:\,m_\tau ~\simeq~ \ep^4\,:\,\ep^1\,:\,\ep^0\,,
\label{eq:LeptonMR}
\eeq
which require,
\beq
(y_1\!-\!u_1)\!-\!(y_3\!-\!u_3) \!=\! \pm 4,~~
(y_2\!-\!u_2)\!-\!(y_3\!-\!u_3) \!=\! \pm 1.
\label{eq:LeptonCond}
\eeq
The tau Yukawa coupling itself can be estimated as 
${\mathrm y}_\tau  \simeq 
(m_\tau/m_t)\tanb\simeq 10^{-2}\tanb\sim \ep^1$
(with $\tanb=\langle H_2\rangle/\langle H_1\rangle$), 
in the typical range of $\tanb \simeq 10-40$, and this restricts the
$U(1)_{\rm FN}$ charges of $\tau$ as $y_3-u_3=\pm 1$. 
Table\,3 summarizes three explicit realizations of Type-II models.   
From Table\,3 and Eq.\,(\ref{eq:ZeeR}), 
the Yukawa couplings of $\nu_R$ are predicted as,
\beq
\ba{ll}
{\rm Type~IIa~\,:}~~  & (f_1,\,f_2,\,f_3) \sim (\ep^3,\,1,\,\ep^1)\,;\\[1.5mm]
{\rm Type~IIb1:}~~    & (f_1,\,f_2,\,f_3) \sim (\ep^5,\,1,\,\ep^3)\,;\\[1.5mm]
{\rm Type~IIb2:}~~    & (f_1,\,f_2,\,f_3) \sim (\ep^{10},\,1,\,\ep^3)\,.
\ea
\label{eq:CoupnuR}
\eeq
From Table\,3 and Eq.\,(\ref{eq:ZeeL}), 
we further predict the left-handed Yukawa couplings $f_{jj'}$,
\beq
\ba{ll}
{\rm Type~IIa~\,:}~~  & 
(f_{12},\,f_{13},\,f_{23}) \sim \ep^{|z|}\,;\\[1.5mm]
{\rm Type~IIb1:}~~    & 
(f_{12},\,f_{13},\,f_{23}) \sim (\ep^{4+z},\,\ep^{6+z},\,\ep^{8+z})\,;\\[1.5mm]
{\rm Type~IIb2:}~~    & 
(f_{12},\,f_{13},\,f_{23}) \sim (\ep^{3+z},\,\ep^{5+z},\,\ep^{12+z})\,;  
\ea
\label{eq:CoupnuL}
\eeq
where the allowed values of $z$ are defined in Table\,3.
Thus, Type-IIa suppresses $f_{jj'}$ couplings to  
$O(10^{-2}-10^{-4})$. 
The models in Type-IIb1 (-IIb2), however,
nicely accommodate the hierarchy (\ref{eq:nuCond}) for the MSW large angle
solution (vacuum oscillation solution), while the predicted size
of $f_{12}\sim 10^{-3}-10^{-6}$ 
is also of the right order\,\cite{Jar}.  Finally, it is
trivial to extend these models with more than one singlet $\nu_{R}$
(i.e., $\nu_{Rj}$ with $j=1,\cdots, N_{\nu_R}$ and $N_{\nu_R}=3$ for instance), 
by simply defining them to share the same $U(1)$ charges as in Tables\,1 and 3.   
\\[2mm]
\noindent
{\small 
Table\,3. 
$U(1)_{\rm FN}$ quantum number assignments for Type-IIa, -IIb1 and -IIb2 models.
[The defined range of $z$ is $|z|\sim 3$ for Type-IIa and 
$0\lesssim z \lesssim 3$ for Type-IIb1 and -IIb2.]
}
\vspace*{-0.6mm}
\begin{center}
\setlength{\tabcolsep}{0.2pc}
\begin{tabular}{c|ccccccccccc}
\hline\hline
&&&&&&&&&&&\\[-2.5mm]
         &  $L_1$  & $L_2$   &  $L_3$    & $e_R$ & $\mu_R$ & $\tau_R$ & $\nu_R$ &
$H_1$    &  $H_2$  & $~S_1^+$ &  $~S_2^+$           \, \\ [1.5mm] 
\hline
&&&&&&&&&&&\\[-2.5mm]
IIa         & $0$        &  $0$  &  $0$  & $-5$       & $-2$    &
$-1$        & $x\!+\!1$  &  $0$  &  $z$  & $-z$       & $1\!-\!x$       \\[1.5mm]
\hline
&&&&&&&&&&&\\[-2.5mm]
IIb1       & $-1$       &  $-3$  &  $-5$  & $4$   & $-1$       &
$-4$       & $x$        &  $0$   &  $z$   & $-z$  & $1\!-\!x$       \\[1.5mm]
\hline
&&&&&&&&&&&\\[-2.5mm]
IIb2       & $2$        &  $-5$  &  $-7$  & $7$   & $-3$       &
$-6$       & $x$        &  $0$   &  $z$   & $-z$  & $3\!-\!x$       \\[1.5mm]
\hline\hline                   
\end{tabular}
\end{center}

\vspace*{4mm}
\noindent
\underline{\it Zee Scalars, Muon $g-2$ and $\mu\to e\gamma$}\\[-2.5mm]

The above minimally extended Zee-type models economically 
incorporate the $\nu_R$ and naturally explain the mass patterns of
the neutrinos and leptons. 
The Zee-scalar Yukawa couplings with the neutrinos/leptons also
exhibit an interesting spectrum.
Now we are ready to analyze their phenomenological impact.
The Brookhaven E821 collaboration has announced a $2.6$
standard deviation in the muon anomalous magnetic moment
$a_\mu = (g_\mu \!-\!2)/2$, i.e.,
$\,
\Delta a_\mu \!\equiv\!
a_\mu^{\rm Exp} \!-\! a_\mu^{\rm {\small SM}} 
\!=\! (42.6\pm 16.5)\times 10^{-10}$\cite{BNL},
which gives a $90\%$\,C.L. range for new physics,
\beqn
15.5 \times 10^{-10} \leq \Delta a_\mu \leq 69.7 \times 10^{-10} \,. 
\label{eq:data}
\eeqn
Different authors\,\cite{others}
have interpreted this anomaly in terms of 
supersymmetry, muon compositeness, extra $Z'$,
leptoquarks and extended neutrino models.
We attempt to explain it from the contribution of
the Zee-scalars and the singlet $\nu_R$ 
in our minimal Type-(I,\,II) models. 

The Zee-scalars $S_1^\pm$ and  $S_2^\pm$  in Type-I/II contribute to 
$g_\mu - 2$ via the Yukawa couplings 
$f_{12,23}$ with $(\mu_L,\,\nu_{e,\tau})$ and  
$f_{2}$ with $(\mu_R,\,\nu_R)$, respectively. 
Thus, we have,
\vspace*{-1mm}
\beqn
\Delta a_\mu &=&
\dis \f{m_\mu^2}{96\pi^2}
     \(\f{|f_{12}|^2+|f_{23}|^2}{\ov{M}_1^2}+\f{|f_2|^2}{M_{S_2}^2}\)
\nonumber\\
           &\simeq& 
\dis
11.8\times 10^{-10} \times
|f_2|^2 \(\f{100\,{\rm GeV}}{M_{S_2}}\)^2,
\label{eq:Sg2}
\eeqn
with
$\ov{M}_1^2 \!=\! (\cos^2\phi/M_{S_1'}^2\!+\!\sin^2\phi/M_{H^{\prime\pm}}^2)^{-2}$.
Here $(M_{S_1'},$ $M_{H^{\prime\pm}})$ are the mass-eigenvalues of the two
charged scalars in Eq.\,(1) and $\phi$ is their mixing angle.  Our models
forbid or highly suppress the mixing between $S_1^\pm$ and $S_2^\pm$
[cf. Table\,2 and Eq.\,(\ref{eq:ZeeCond})].
Note that the $f_{12,23}$ terms in (\ref{eq:Sg2})
are negligible compared to the $f_2$ term for Type-I and 
Type-II (with $|z|\geq 1$), cf. Eqs.\,(\ref{eq:CoupnuR})-(\ref{eq:CoupnuL}). 
The precision bound from $\mu\to\nu_\mu e\ov{\nu}_e$ decay gives\,\cite{Jar},
$f_{12}/\ov{M_1}< 0.18/$TeV, which, combined with 
Eq.\,(\ref{eq:nuCond}), also renders
the $f_{12,23}$ terms irrelevant for the $\gm2$ anomaly. 
Hence, the original Zee-model {\it cannot}  resolve the $\gm2$ anomaly.
From (\ref{eq:data}) and (\ref{eq:Sg2}),
we deduce,
\vspace*{-1.5mm}
\beq
41.1\,{\rm GeV} \leq \dis\f{M_{S_2}}{|f_2|} \leq 87.2\,{\rm GeV} \,.
\label{eq:boundMSf2}
\eeq
Since the LEP2 direct search for charged particles requires
$M_{S_2} \gtrsim 100$\,GeV, $|f_2|$ is constrained as,
\vspace*{-1mm}
\beqn
1.1\lesssim |f_2| \lesssim O(1)\,,
\label{eq:f2bound}
\eeqn
where the upper bound is imposed by perturbativity.  
Eq.\,(\ref{eq:f2bound}) is in the predicted range of $f_2$ for our Type-II
models [cf. Eq.\,(\ref{eq:CoupnuR})]. 
Thus, combining (\ref{eq:boundMSf2}) and (\ref{eq:f2bound}), we
conclude that, to fully accommodate the BNL $\gm2$ data, 
the Zee-scalar $S_2^\pm$ in 
our Type-I and -II models has to be generically light, around $100-300$\,GeV.
This leads to the possibility of discovering the
light charged Zee-scalar at the Tevatron Run-2,
the LHC or a future high energy Linear Collider.
Since $f_2 \gg f_{1,3}$ for Type-II,
the  $S_2^\pm \to \mu_R^\pm\nu_R$ decay has 
a large branching ratio, and
dominates over the $e_R^\pm\nu_R$ and $\tau_R^\pm\nu_R$ modes.
Though Type-I models do not predict $f_{1,3}$, 
they allow $f_2=O(1)$
while $f_{1,3}\ll f_2$ is forced by 
the $\mu\to e\gamma$ and $\tau\to \mu\gamma$ bounds below.
We have the partial decay width,
\vspace*{-2mm}
\beqn
\Gamma [S_2^\pm \to \mu^\pm_R\nu_R] &\simeq& 
\dis \f{f_2}{16\pi^2}M_{S_2}\simeq O(1)\,{\rm GeV}, 
\eeqn
for $M_{S_2} \simeq 100-300$\,GeV.
Hence, $S_2^\pm$ is a very narrow spin-0 resonance.
The predicted branching ratio 
${\rm Br}[S_2^\pm \to\mu^\pm_R\nu_R]\simeq 1$ 
in Type-(I,\,II) 
suggests that $S_2^\pm$ can be best detected via 
muon plus missing energy.

Our models  have further implications for the
flavor-violating rare decay $\mu\to e\gamma$, whose
branching ratio is bounded by,
${\rm Br}[\mu\to e\gamma] < 1.2\times 10^{-11}$,
at the $90\%$\,C.L.  
The Type-(I,\,II) models give the following contributions,
\vspace*{-2mm}
\beqn
{\rm Br}[\mu\to e\gamma] &=& \dis
\f{\alpha_{\rm em}v^4}{384\pi} 
\(\f{|f_1f_2|^2}{M_{S_2}^4}+
  \f{|f_{13}f_{23}|^2}{\ov{M}_1^4}\)  \nonumber\\
&\simeq&
\f{\alpha_{\rm em}}{384\pi} 
 |f_1f_2|^2\({v}/{M_{S_2}}\)^4 \,,
\eeqn
where $v=(\sqrt{2}G_F)^{-1/2}\simeq 246\,$GeV.
Thus, we derive,
\vspace*{-1.5mm} 
\beqn
|f_1f_2|< 2.3\times 10^{-4}\({M_{S_2}}/{100\,{\rm GeV}}\)^2 \,.
\eeqn
Combining this with the $\gm2$ bound in (\ref{eq:boundMSf2}),
we find,
\vspace*{-1.5mm} 
\beq
\left|{f_1}/{f_2}\right| < 1.8\times 10^{-4} \,,
~~ {\Rightarrow} ~~ f_1 \lesssim (2-6)\times 10^{-4}\,.
\label{eq:muega}
\eeq
Comparing this with Eq.\,(\ref{eq:CoupnuR}), we 
see that Type-IIb1 has $f_1$ just below the current bound while Type-IIb2
is well below it. On the other hand, the $f_1$ coupling in Type-IIa lies 
slightly above the limit by a factor of $2-3$;  given the
uncertainty of the parameters, it can be easily adjusted to stay
within the bound. 
Also, a much weaker bound on $f_3$ can be derived from
$\tau\to\mu\gamma$ decay, i.e.,
$ |f_3| \lesssim 0.06-0.16 \sim O(0.1) $ at $90\%$\,C.L.,  
for $1.1 \lesssim |f_2| \lesssim 3$,
which is consistent with the Type-II predictions in (\ref{eq:CoupnuR}).
Finally, if we include $N_{\nu_R}(\geq 2)$ singlet $\nu_{Rj}$  
with the same Yukawa coupling $f_2$,
the upper [lower] bound in Eqs.\,(\ref{eq:boundMSf2}) and  (\ref{eq:muega})
[Eq.\,(\ref{eq:f2bound})] 
will be relaxed by a factor of $\sqrt{N_{\nu_R}}$.

\vspace*{0.2cm}
In summary, the Zee model naturally generates small neutrino Majorana masses by 
radiative corrections, but it neither predicts the Zee-scalar Yukawa couplings
nor provides any insight on the lepton mass hierarchy.
We have constructed a class of minimally extended Zee-models with the right-handed
neutrino $\nu_R$ embedded, where a $U(1)$ symmetry protects the radiative
neutrino masses while generating the lepton mass hierarchy, 
the hierarchy of the Zee-scalar 
Yukawa couplings required by the neutrino oscillation data, 
the hierarchy of Zee-scalar Yukawa
couplings necessary for consistency with the $\mu\to e\gamma$ bound, 
and the size of the Zee-scalar Yukawa coupling 
needed for the BNL $\gm2$ anomaly. 
Furthermore, a light Zee scalar $S_2^\pm$ 
is predicted in our models, with a mass around $100-300$\,GeV.

\vspace*{2.6mm}
This research was supported in part by the U.S. Department of Energy 
under Contract No. DE-FG03-93ER 40757 and 
by the Natural Science and Engineering Research Council of Canada.
We thank Sasha Kopp and Ernest Ma for discussions.

\end{document}